# Effects of architectural issues on a km$^3$ scale detector

R. Coniglione[a] for the NEMO collaboration[*]

*Laboratori Nazionali del Sud – INFN, Via S. Sofia 62, Catania 95123 Italy*

**Abstract**

Simulation results showing the comparison between the performance of different km$^3$ detector geometries are reported. Effective neutrino areas and angular resolutions are reported for three different geometries based on "NEMO-towers" and strings. The results show that the NEMO-tower based detector has the best performance concerning both the effective area and the angular resolution isotropy.

*Keywords:* Type your keywords here, separated by semicolons ;

## 1. Introduction

One of the main goals of the astroparticle physics is the detection of high energy neutrino fluxes from galactic and extragalactic sources. The theoretically predicted neutrino fluxes indicate that detectors with a detection area of about 1 km$^2$, located in low background environments, should be able to detect HE neutrino fluxes. Collaborations, that aim to construct large-scale Cherenkov detectors, are already at work. In particular, the IceCube collaboration [1], that operates in the South Pole, is installing a km$^3$ detector based on about 4800 PMT sensors arranged in 80 vertical strings deployed at a maximum depth of 2400 m in the ice. In Europe, an international consortium (KM3Net) [2] is already active for a design study of a km$^3$ detector to be installed in the Mediterranean sea. This collaboration takes profit of the know-how of the ANTARES [3], NESTOR [4] and NEMO [5] collaborations that are already active in this field since many years. The two telescopes, installed in the two Earth hemispheres, will assure full sky coverage and the detection of transient phenomena. In particular from the Northern hemisphere the galactic center region, where very active TeV gamma sources have been recently observed by HESS [6], is observable.

---

[*] Corresponding author. Tel.: +39095542288; fax: +390957141815; e-mail: coniglione@lns.infn.it.

Since 1998 the NEMO collaboration has undertaken an R&D activity for the construction of a km$^3$ detector in a site close to the italian coasts. With this aim the collaboration has explored for long time, performing many sea campaigns, the water and the environmental properties of several sites and has elected as adapt to host a km$^3$ neutrino telescope, a site near the southernmost cape of Sicily (Capo Passero) at a depth of 3400 m. A technological survey to investigate the project feasibility and the realization of prototypes to validate the technological feasibility are also been undertaken by the collaboration.

During these years, all the NEMO activities has been supported by numerical simulations. In this paper we will present some results about estimation of the performance of km$^3$ detectors based on different "structures" that support the PMT, namely "NEMO-towers" [7] and strings.

**2. The simulations**

The design of a km$^3$ detector that has to be installed in a hostile environment as the deep sea is a very hard task. It is the result of a compromise between the technical feasibility and the performance. Of course the total cost has to be taken into account.

One of the main component in the design of a km$^3$ detector is the structure that host the PMTs. It has to fulfill the main following requirement:

- it should be easy to deploy and recover in a single sea operation.
- it should host a large number of PMT in order to minimize the number of structures to deploy.
- it should be mechanically stable with respect to the sea currents

The NEMO collaboration, taking into account these requirements, has proposed a three-dimensional structure called "NEMO tower" [7]. It consists of a sequence of bars interlinked by a series of ropes. Each bar is hortogonal with respect to the adjacent one and host 4 PMT's, two (one down-looking and the other horizontally looking) at each bar edge. This PMT configuration permits to apply two-fold or three-fold local coincidences between PMTs in the same bar. This local coincidences strongly suppress hits from background noise.

An alternative kind of structures, already adopted by the IceCube and ANTARES collaborations, is the string, namely, a series of PMTs, which may also be clusters composed by several PMTs, arranged along a line. A full R&D activity for a km$^3$ detector based on string has not yet started.

The detector sensitivity of a neutrino telescope to neutrino fluxes from different astrophysical sources depends on the neutrino effective areas for up-going neutrinos and on the reconstruction capability of the neutrino direction.

The software package used for the simulations presented in this work is the software developed by the ANTARES [8] collaboration modified for a km$^3$ detector [9]. We simulated an up-going muon neutrino flux with a spectral index X=2 in the energy range between $10^2$ GeV and $10^8$ GeV. Effective areas for three different detector lay-outs were simulated: a km$^3$ geometry made of NEMO-towers (NEMOdh-

| Detector configuration | NEMOdh-140 | RETd-125 | RETdh-14 |
|---|---|---|---|
| N° of PMTs | 5832 | 5800 | 5832 |
| N° of floors | 18 | 58 | 18 |
| N° of structures | 81 | 100 | 81 |
| Structure distance | 140 m | 125 m | 140 m |
| Structure height | 680 m | 912 m | 680 m |
| Bar length | 20 m | | |
| Floor distance | 40 m | 16 m | 20 m |
| Instrumented volume | 0.88 km$^3$ | 1.15 km$^3$ | 0.88 km$^3$ |
| PMT diameters | 10" | 10" | 10" |
| PMT configurations | Down-horizontal | Down | Down-horizontal |

Table 1. Main detector parameters of the simulated geometries

140) and two geometries made of strings, RETd-125 with one down-looking detector each 16 m and RETdh-140 with two PMTs, down and horizontally looking, each 20m. For all the geometries examined the structures were arranged in a square pattern. The main lay-out parameters are reported in Table 1.

In the simulation the environmental parameters measured in Capo Passero [10] have been taken into account. In particular an absorption length of 70 m at 440 nm and a $^{40}$K background rate of 30 kHz have been considered. In order to have a fair comparison between the effective areas for the three geometries, quality cuts on the reconstruction parameters [11] have been applied. In Fig.1 top panel the median of $\Delta\Omega$ distribution, the angle between the neutrino direction and the reconstructed muon direction, is shown. Cuts were chosen in order to reach for each geometry an angular resolution at around 1 TeV, close to the angle between the neutrino and the muon directions at the interaction point. The correspondent effective areas as a function of the neutrino energy are reported in Fig. 1 central panel. The value of effective areas reaches a value of around 1 km$^2$ at 10 TeV for all the three geometries. At lower energies, where the PMT lay-out plays the major role, the NEMOdh-140 shows a significantly higher value. At the highest energies the instrumented volume is much more important. In the bottom panel of Fig. 1 the ratio between the effective areas for the NEMOdh-140 and the RETdh-140, the two geometries with the same main detector parameters (N° of PMTs and structures, instrumented volume, structure height..) is reported.

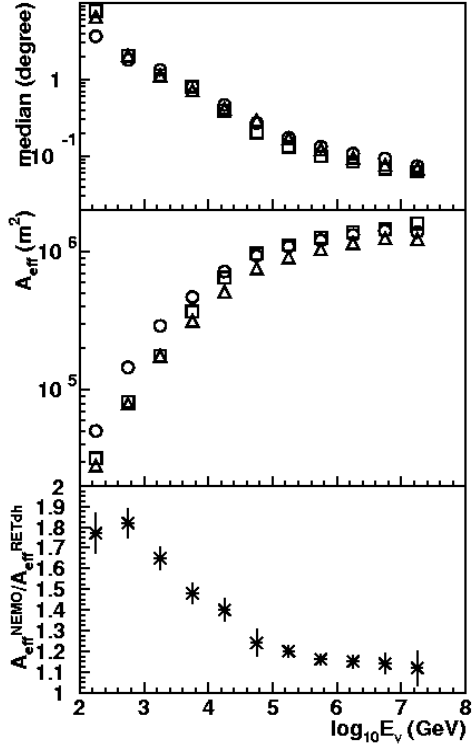

Fig. 1 - Median (top panel) of the distribution of the angle between the neutrino direction and the reconstructed muon direction and the effective area (central panel) as a function of the neutrino energy for the NEMOdh-140 (circles), the RETd-125 (squares) and the RETdh-120 (triangles). In the bottom panel the ratio between the NEMOdh-140 e RETdh-140 lay-out effective areas is reported.

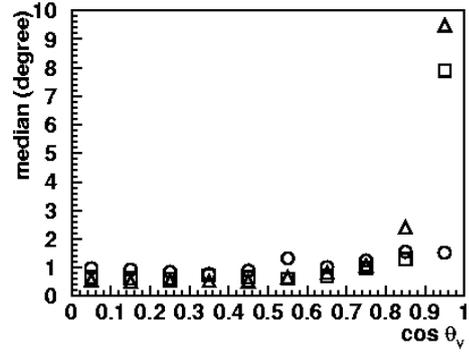

Fig. 2 - Median of $\Delta\Omega$ as a function of the neutrino direction (cos $\theta_v$ = 0 orizontal neutrinos, cos $\theta_v$ =1 up-going vertical neutrinos) for neutrino energy 1TeV $\leq$ E$_v$ $\leq$ 10 TeV for the three geometries simulated (same symbols of Fig. 1).

In order to evaluate the isotropy of the reconstruction capability of the telescope, in Fig. 2 the median of the $\Delta\Omega$ distribution as a function of the incoming neutrino direction (1 TeV $\leq$ E$_v$ $\leq$ 10 TeV) is reported for the simulated geometries. For the geometries based on strings a worse reconstruction for near vertical neutrinos is observed and this trend is weaker and weaker with increasing the neutrino energy.

In this work the performance of km$^3$ detector based on a three-dimensional structure, the "NEMO-tower", and strings has been explored simulating the response of the detector for up-going neutrino. The geometry based on NEMO-tower compared with detectors based on strings shows the best performance both in terms of effective areas and on angular homogeneity of the reconstruction capability. Futher simulations, that include the atmospheric neutrino and muons backgrounds, are necessary to evaluate the detector sensitivity.